%% file: main.tex
\documentclass[sigconf, acmlarge]{acmart}

\AtBeginDocument{%
  \providecommand\BibTeX{{%
    \normalfont B\kern-0.5em{\scshape i\kern-0.25em b}\kern-0.8em\TeX}}}

\copyrightyear{2022}
\acmYear{2022}
\setcopyright{rightsretained}
\acmConference[FAccT '22]{2022 ACM Conference on Fairness, Accountability, and Transparency}{June 21--24, 2022}{Seoul, Republic of Korea}
\acmBooktitle{2022 ACM Conference on Fairness, Accountability, and Transparency (FAccT '22), June 21--24, 2022, Seoul, Republic of Korea}\acmDOI{10.1145/3531146.3533193}
\acmISBN{978-1-4503-9352-2/22/06}
\acmDOI{10.1145/3531146.3533193}

\usepackage{xcolor}
\usepackage{dsfont}
\usepackage{multirow}
\usepackage{arydshln}
 \usepackage{mdframed}
 \usepackage{todonotes}
 \mdfdefinestyle{MyFrame}{
    outermargin=5pt,
    innerrightmargin=3pt,
    innerleftmargin=3pt,
    innertopmargin=4pt,
    innerbottommargin=4pt
}
\usepackage{courier}
\newcommand{\pname}{Exp-HAIC}

\begin{document}

\title[Who Goes First?]{Who Goes First? Influences of Human-AI Workflow\\ on Decision Making in Clinical Imaging}

\author{Riccardo Fogliato}
\email{rfogliat@andrew.cmu.edu}
\affiliation{%
  \institution{Carnegie Mellon University}
  \country{USA}
}

\author{Shreya Chappidi}
\affiliation{
  \institution{University of Virginia}
  \country{USA}}

\author{Matthew Lungren}
\affiliation{
\institution{Stanford University}
\country{USA}}

\author{Michael Fitzke}
\author{Mark Parkinson}
\affiliation{%
  \institution{Mars Digital Technologies}
  \country{USA}
}

\author{Diane Wilson}
\author{Paul Fisher}
\affiliation{
\institution{Antech Imaging Services}
  \country{USA}}

\author{Eric Horvitz}
\author{Kori Inkpen}
\author{Besmira Nushi}
\affiliation{%
 \institution{Microsoft Research}
 \country{USA}}

\renewcommand{\shortauthors}{R. Fogliato et al.}

\begin{abstract}
Details of the designs and mechanisms in support of human-AI collaboration must be considered in the real-world fielding of AI technologies. 
A critical aspect
of interaction design for AI-assisted human decision making are policies about the display and sequencing of
AI inferences within larger decision-making workflows.
We have a poor understanding of the influences of
making AI inferences available before versus after human review of a diagnostic task at hand.
We explore the effects 
of providing AI assistance at the start of a diagnostic session in radiology
versus after the radiologist has made a provisional decision. 
We conducted a user study
where 19 veterinary radiologists 
identified radiographic findings present in patients' X-ray images, 
with the aid of an AI tool. 
We employed two workflow
configurations to analyze 
(i) anchoring effects, 
(ii) human-AI team diagnostic performance 
and agreement, 
(iii) time spent and confidence in decision making, and 
(iv) perceived usefulness of the AI.
We found that participants who are
asked to register provisional responses 
in advance of reviewing AI inferences 
are less likely to agree with the AI 
regardless of whether the advice is accurate and, in instances of disagreement with the AI, 
are less likely to seek the second opinion of a colleague. 
These participants also reported the AI advice to be less useful. 
Surprisingly,
requiring provisional decisions on cases in advance of the display of AI inferences  did not lengthen the time 
participants spent on the task. 
The study provides generalizable and actionable insights
for the deployment of clinical AI tools in human-in-the-loop systems and introduces a methodology for studying alternative designs for human-AI collaboration. 
We make our experimental platform available 
as open source to facilitate future research on the influence of alternate designs on human-AI workflows.
\end{abstract}




\maketitle


\input{text/introduction}

\input{text/background}

\input{text/experiment}

\input{text/results}
\input{text/discussion}

\input{text/conclusions}

\bibliographystyle{ACM-Reference-Format}
\bibliography{references}
    

\end{document}

%% file: text/introduction.tex

\section{Introduction}

We explore the influences of the sequencing of the availability of AI inferences on human decision making in a clinical imaging setting. We assess whether eliciting initial diagnoses from the participant before revealing the AI recommendation influences their final decisions and overall usage of AI inferences. In the study, we experiment with two human-AI collaboration workflow configurations. In the {\it one-step workflow}, participants were asked to identify radiographic findings given AI inferences and X-ray images at the same time. In the {\it two-step workflow}, participants were presented with AI inferences only after they had made a provisional decision. The two workflows reflect distinct approaches to interleaving AI assistance with human decision making. In practice, the joint presentation of the radiographic findings and inferential analysis at the start provides a more comprehensive set of information sources to the decision maker. However, as AI inferences may be erroneous, there are concerns about the possibility of unwanted anchoring that could lower the team's performance \cite{bansal2021does, buccinca2021trust, fogliato2021impact, rastogi2020deciding}. 

Beyond focusing on performance, deployments must also consider preferences of human decision makers about alternate workflows when it comes to usability and adoption. While there is enthusiasm about bringing AI inferences into practice, resistance has been noted to AI assistance, with basis in multiple factors, including changes in established patterns of practice and aversion to automation \citep{dietvorst2015algorithm,burton2020systematic,gaube2021ai}. 
For example, in high-stakes domains where human decision makers are the ultimate decision makers, there may be concerns regarding the influence of AI assistance on predictive performance, effort, and productivity. To understand the multifaceted impact of these tools, we measured (i) anchoring effects, (ii) human-AI team diagnostic performance and agreement, (iii) time spent and confidence in decision making, and (iv) perceived usefulness
of the AI inferences.

We examined the influence of alternate human-AI workflows on a clinical imaging task with 19 veterinarian radiologist participants based at Mars, a pet healthcare company. The radiologists were asked to inspect and identify 33 different findings in X-ray images from real-world cases that had come to the company.
In both workflows, the AI assistance consisted of binary AI inferences (finding present versus absent) on each finding along with the respective confidence scores.
AI inferences were obtained from an ensemble machine learning model that is under consideration by Mars for deployment. 
The study was conducted via a web-based experimentation platform (Figure~\ref{fig:interface}) where participants could inspect X-ray images, register their findings, and review AI inferences.

Key findings from the study demonstrate that alignment between participants' diagnoses and AI inferences is strongest in the one-step workflow, where radiologists were presented with AI assistance at the beginning of the diagnostic sessions. 
The findings highlight a higher risk of anchoring in the hone-step AI workflow. 
While we had hypothesized an anchoring effect given the nature of the workflow, we found that anchoring effects were minimal for findings considered critical or life-threatening for the animal. 
Although the AI outperformed participants, anchoring effects only led to marginal gains in diagnostic performance due to over reliance on erroneous AI advice.
From a productivity perspective, we found that the time spent in both workflows was comparable. We were surprised to found that participants in the two-step workflow rarely revised their provisional diagnoses when the AI inferences differed from their earlier assessment. 
In perceptions shared in a survey, participants in the one-step workflow expressed a sense that the AI increased their confidence and speed more than those in the two-step workflow, and rated AI inferences as more useful. We believe that our multi-dimensional analysis provides actionable insights on the fielding of AI assistance in clinical imaging domains, suggesting that automated inferences  may be most beneficial to human decision making when it is least disruptive per being smoothly integrated into the flow of human cognitive processes.

In summary, we make the following contributions:
\begin{itemize}
    \item  We conduct a user study to investigate the influence of two human-AI workflows on the diagnoses made by expert veterinary radiologists with the aid of an AI diagnostic tool. The analysis investigates   questions about the influence of the sequencing of AI inferences on key dimensions, including anchoring bias, diagnostic performance, agreement, time spent, and user satisfaction.
    \item Based on the study results, we derive and discuss a set of implementable takeaways for the deployment of AI tools in human-driven decision-making processes. 
    \item We release as open source the experimental platform, which can be used for conducting human-in-the-loop user studies in clinical imaging. The code is available at  \texttt{\url{http://aka.ms/\pname}}. 
\end{itemize}

The rest of the paper is organized as follows. In Section~\ref{sec:background}, we position and contrast our work with previous findings in the human-AI collaboration and decision-making literature. 
Section~\ref{sec:methods} contains details of the experimental setup and the platform. 
Section~\ref{sec:results} describes the study findings.
Sections ~\ref{sec:newdiscussion} and ~\ref{sec:newconclusion} discuss takeaways and future research directions.


%% file: text/background.tex
\section{Background}
\label{sec:background}

\subsection{Analyses of Human-AI Teams}

AI tools are being deployed to aid human decision makers 
in a variety of high-stakes domains 
including healthcare, criminal justice, child welfare, and hiring
\citep{desmarais2020predictive, de2020case, raghavan2020mitigating}. In medicine, 
the recent advent of deep learning methods 
has sparked enthusiasm about translating prototypes in practice
\citep{topol2019high, futoma2017learning, ribli2018detecting, haque2020deep},
with systems showing performance on par with experts on diagnostic tasks \citep{wu2019deep,
esteva2017dermatologist, haenssle2018man, majkowska2020chest,
tschandl2019comparison, makimoto2020performance, rajpurkar2017chexnet,
maron2019systematic, hekler2019deep, brinker2019convolutional}.
The hope is that 
these tools will produce sizable gains in efficiency of human decision-making processes 
\citep{mckinney2020international, litjens2016deep, ruamviboonsuk2019deep,
bayati2014,Wiens2016, lee2020icu, kleinberg2018human}. 
However, evidence to date suggests that their deployment of AI systems does not necessarily yield 
a uniform improvement over the status quo \citep{stevenson2018assessing, skeem2020impact, beede2020human, esmaeilzadeh2015adoption}.

The adoption of these tools 
has fostered a scholarly effort 
on designing human-AI collaborations for optimal team decision making
\citep{madras2017predict, bansal2021does, hilgard2021learning, wilder2020}. 
Past studies in this space 
analyze how user performance and trust are affected 
by the presence of AI explanations \citep{zhang2020effect, poursabzi2021manipulating, lai2019human, bansal2021does}, 
the perceived and communicated AI accuracy
\cite{yin2019understanding,lai2019human},
and model updates \cite{bansal2019updates}, 
among others. 
A common theme in this body of work is that, 
when building an AI intended to collaborate with a human, 
numerous details of human-centered design
need to be considered, in contrast to the dominant focus of
attention on AI accuracy. 
Our study contributes new insights about the importance of workflow configurations 
as part of designs for human-AI collaboration. 
In contrast to prior studies, we analyze the decision making of domain experts, rather than of laypersons, on a  diagnostic task in the medical space.

Prior studies have also analyzed human-AI teams
in the clinical imaging setting. 
The studies to date largely focus on a comparison of diagnostic performance 
of humans alone versus human-AI teams
\citep{steiner2018impact,han2020augmented,taylor2008ct,
bien2018deep,jain2021development, wang2021deep, lehman2015diagnostic}. 
These analyses, 
for the most part, 
report that interactions with AI tools lead to gains in human diagnostic performance. 
A handful of studies have compared the influences 
of various types of AI assistance on decisions \citep{sayres2019using,tschandl2020human, cai2019human}. 
Results from these investigations indicate that
AI tools appropriately designed to support decision makers 
can boost not only diagnostic accuracy 
but also self-reported confidence, 
while decreasing mental effort and the time spent on the task. 
At the same time, 
some of these analyses have also witnessed the pitfalls of AI adoption, including
increases in the time spent on the task without corresponding gains in accuracy, 
reductions in diagnostic performance due to reliance on erroneous AI advice, 
and participants ignoring AI advice altogether
\citep{tschandl2020human,jain2021development, lehman2015diagnostic}. 
These are notable limitations because 
the performance of machine-learned models often varies across types of instances 
and can degrade over time \citep{tschandl2019comparison, degrave2020ai}. Poor integration of AI in human decision-making workflows can hamper the adoption of the tool by real-world decision makers \citep{lugtenberg2015implementation, khairat2018reasons}. 
Although our experiment cannot fully emulate the clinical setting in which radiologists operate, our analysis attempts to capture the impact of the workflows across various dimensions.

\subsection{Workflow Considerations for Human-AI Teams} 

A critical aspect of how AI can influence decision making revolves around
the bias of anchoring \citep{Tversky_Kahneman_1974}. Multiple studies have demonstrated that
people may give stronger weight to their assessments towards prior knowledge 
or analyses 
versus doing full revisions in light of new evidence 
\cite{wang2019designing}. 
We hypothesized that anchoring effects of the review of information would be stronger when presented early versus late in problem solving. Thus we expected that AI inferences presented at
the same time as initial analysis would be more influential than when the inferences are presented after an initial assessment. 
Research on the explanation of AI inferences frames opportunities for further study of the influences of designs for workflow of human-AI collaboration, including altering the timing of AI-assistance 
and forcing users to spend more time on instances where AI inferences present higher uncertainty 
\cite{bansal2021does, buccinca2021trust, 
rastogi2020deciding, gajos2022people, park2019slow}. 

Several studies on human-AI collaboration have focused specifically on anchoring and workflow orderings similar to those studied in our experiment. 
\citet{green2019principles} find that requiring participants to register provisional predictions 
before AI recommendations are revealed 
results in marginal gains in overall predictive performance. 
In a similar experimental setup, 
\citet{fogliato2021impact} do not detect anchoring effects or differences in performance across workflows. 
\citet{buccinca2021trust}
report lower reliance on erroneous AI advice in the two-step workflow; we find similar results. 
In distinction to the these prior studies, rather than studies with lay participants,we study the behaviors and perceptions of domain experts on the tasks they perform in the course of their professional work. 
Assessments such as ours are critical requirements for real-world deployments because experts may have deeply ingrained processes for decision making and thus may interact with AI tools differently from crowdworkers employed in most studies of human-AI interaction. 
For example, experts may be reluctant about reviewing and leveraging AI advice \citep{gaube2021ai, cheng2022heterogeneity}, a phenomenon referred to as ``algorithm aversion'' \citep{dietvorst2015algorithm}.

Beyond anchoring effects, we need to consider the cognitive cost of different sequencing of information fusion and decision making. Psychologists have shown that decision makers seek to minimize cognitive effort based on considerations of the perceived costs and benefits of the mental effort associated with different strategies for coming to a decision \cite{Kooletal2010}. In this realm of research, studies have identified challenges with cognitive costs of aggregating new evidence \cite{BroderSchiffer2003, Payne1988} and with considering sets of alternatives \cite{HauserWernerfelt1990}. The cognitive effort required in a two-step versus one-step workflow to consider new information and to re-evaluate prior assessments has conceptual links to studies on the costs associated with task switching, interruption, and recovery \cite{CCHtaskswitch2000,iqbal2007disruption,hkph2003cacm}. Cognitive costs of re-examination when new information becomes available can be viewed as analogous to interruption and recovery on the initial task with new information \cite{HACostMultiTask2000,hka2004cscw}. Thus, the re-opening of a completed analysis, as required in a two-step workflow, will tend to increase the cognitive effort required for a decision. 

In findings related to cognitive effort, research on ``cognitive forcing'' has explored methods for pushing human decision makers to spend more time with deliberating about problems \citep{buccinca2021trust, rastogi2020deciding, park2019slow, gajos2022people}. Work in this area includes making AI assistance only available upon request or employing a "slow algorithm" that loads while the user waits to input their decision. While these cognitive forcing functions were found to increase performance measures and decrease AI reliance, they did so at the expense of additional time required for decision making \cite{shane2015reflection}. Findings from other studies indicate that it is difficult for humans to revise or reverse their decisions due to psychological phenomena of sunk cost effects \cite{Domeierer2018motivational, ArkesBlumer1985}, cognitive dissonance \cite{Festinger1959}, and confirmation bias \citep{Klayman1995, Nickerson1998}. Moreover, \citet{Kirkeboen2013} find that decision reversals are associated with higher levels of post-outcome regret despite improved outcomes or predictive performance. 


%% file: text/experiment.tex

\section{Methods}
\label{sec:methods}

We now describe task, experimental design, procedures,  measures, data analysis, and experimental platform. 
The study received IRB approval by Mars. All X-ray images in the study were drawn from past patient examinations. 

\subsection{Experimental Task}\label{sec:task}

\paragraph{Task} Study participants
were shown 40 X-ray images
from individual veterinary patients, 
which were all dogs for study consistency. 
The images were divided 
into two series of 20 images 
to reduce the load of a single session.
For each image, 
radiologists were asked to diagnose 
which of 33 pre-specified radiographic findings
could be identified. 
As shown in Figure~\ref{fig:interface}, 
the diagnostic task for each of the findings required the following:
\begin{itemize}
\item Estimating the likelihood that the finding was present in the X-ray.
\item Assessing whether to call the finding as present or absent in the X-ray.
\item Flagging whether a second opinion from a colleague was needed. 
\end{itemize}
The questions were modeled after assessments that radiologists make in their daily jobs.
Response to the first request to assess the likelihood of findings
were provided via a 0\%--100\% slider
 with bins of 10\%.
The remaining assessments were input using
yes/no radio buttons. 
All answers were initialized to $0\%$ 
and ``no'' respectively. 

When making diagnoses, 
participants were assisted by an AI diagnostic tool 
that estimated the likelihood of each finding
being present in the X-ray image. See  \citet{fitzke2021rapidread} for a detailed description of the tool.
The estimates are an average of predictions
generated by eight separate convolutional neural networks, 
trained on a large proprietary dataset. 
To convert the estimates produced by the AI tool into binary predictions, 
we adopted a threshold of 0.6 
which corresponds to the choice made 
while deploying the model in production to help radiologists label new data for AI training purposes. 
This threshold was derived by calibrating each model to maximize its probability predictions at 0.5 with regards to Youden's J-Statistic \citep{youden1950index} and then further calibrating the resulting ensemble to 0.6 due to its accuracy gains \citep{fitzke2021rapidread}.
Participants had access to both the likelihood and the binary prediction of present versus absent generated by the AI tool for each finding, 
which in the remainder of the paper 
we call {\it AI confidence} and {\it AI flags} respectively.

\paragraph{Dataset}
The X-ray images were obtained from a benchmark dataset previously annotated by 10 to 13 expert radiologists independently. The radiologists labeled the 33 radiographic findings.
We constructed ground truth labels based on these diagnoses by
considering the finding as present 
when at least half of the radiologists had identified it in the image. 
For our study, we obtained 40 X-ray images from this dataset
by oversampling those 
with the lowest agreement among radiologists to boost the difficulty of the task 
and thus maximize the power of our analysis on anchoring effects. Fleiss' kappa, a measure of inter-rater reliability, in the ground truth annotations was 0.44, which indicates weak agreement among radiologists \citep{mchugh2012interrater}. 
In our final sample, only about 7\% of all findings were present according to the ground truth majority vote annotations (91 out of 40$\times$33=1320).

\subsection{Experimental Design}

\paragraph{Treatments} To test the impact of workflow configurations on decision making, 
we employed a between-subjects design 
by assigning each of the participants 
to one of two
workflow configurations. 
In the {\it one-step workflow}, 
the X-ray image 
and the AI inferences 
(AI confidence and binary estimate) 
were shown at the same time. 
In the {\it two-step workflow},
participants were asked 
to make provisional diagnoses 
before the AI inferences
were revealed. 
After seeing the automated inferences, 
they were allowed to revise
their initial diagnoses.
The studied workflows represent easily implementable 
human-AI team configurations that the company is considering for deployment. 
All participants reviewed 
the same two series of 20 images.  
Within each series, the images were reshuffled 
in random order for every participant 
to avoid ordering effects.

\begin{figure*}
    \begin{center}
    \includegraphics[width=\textwidth]{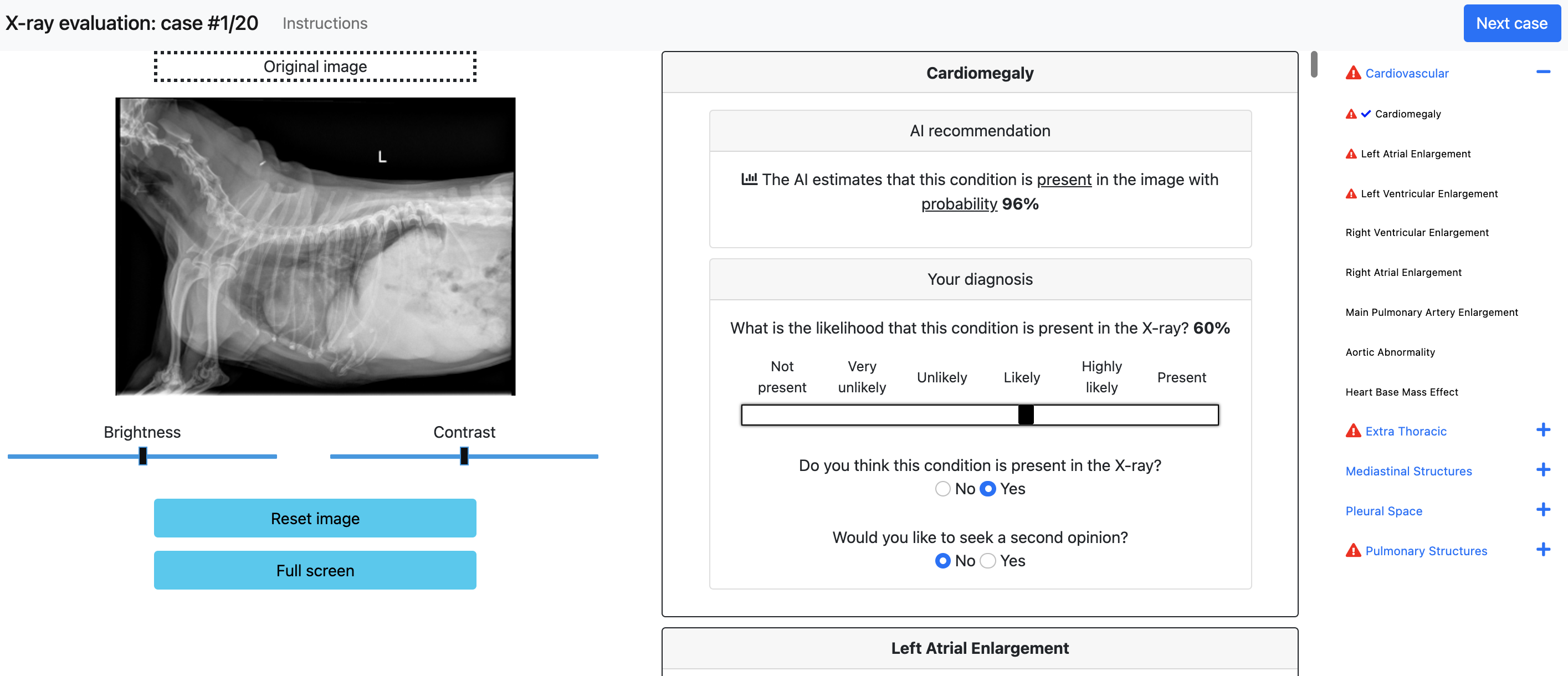}
    \end{center}
    \caption{Screenshot of the interface. On the left, an X-ray image of the thorax and abdomen of a dog is displayed.  
    In the middle, the names of the radiographic findings with respective AI confidence are displayed in each of the boxes. Participants are asked to estimate the likelihood of the finding being present, whether to diagnose it as present or not, and whether they would seek a second opinion (default values are ``0\%'', ``No'', and ``No'' respectively). Findings are grouped into macro-categories in the navigation bar on the right. When the AI flags a finding, a red triangle appears next to the name of both the finding and of the corresponding macro-category. If the participant identifies the finding as present, a check mark is shown.}
    \label{fig:interface}
\end{figure*}


\paragraph{Procedure}

We now describe how participants navigated through the web-based experimental platform. 
Participants were first shown a consent form and asked to provide
an identifier they had been assigned in order to preserve anonymity. 
Next, they followed a series of instructions that included information about the content of the task 
and a description of the AI diagnostic tool. 
Importantly, we clarified in layman's terms that 
the AI confidence may not reflect frequentist probabilities. 
At the end of the instructions, participants
completed a screening test with 10 questions 
designed to ensure that they
understood the task structure
and the information provided by the AI. Participants were prompted to revise their responses until they answered all questions correctly.  
After taking the screening test, participants reviewed each of the 40 images,
one at a time, in two 20-image sessions, using the interface shown in Figure \ref{fig:interface}.
The X-ray image was shown on the left of the web page and participants could zoom in and out, change brightness and contrast, or enlarge it to full-screen. These operations are typically available during X-ray evaluations.
Diagnoses for each of the 33 radiographic findings could be made using the UI controls within the stacked frames in the middle of the web page, which also contained the corresponding AI confidence.
To help participants easily navigate through the findings, a navigation bar appeared on the right of the page. Red triangles were shown next to the names of the findings flagged by the AI, i.e., those with AI confidence$\geq$60\%. 
Check mark symbols were displayed next to the findings identified by the participant in the image. 
AI confidence and flags were hidden from two-step workflow participants during their initial review of the image, and made visible only after they clicked a button in the top right corner of the interface. 
At the end of each image review, participants were asked whether the AI help had been useful, and then could proceed to the next image. 
Participants could not skip images or change diagnoses previously made. 
At the end of the experiment, they were asked to complete a
questionnaire that we discuss in detail in Section \ref{sec:measures}. 


\paragraph{Participants} 
A total of 24 veterinary
radiologists employed at Mars were initially selected to participate in
the study. Half of these radiologists were involved in data labeling of X-ray imagery as part of an ongoing organizational effort to embed machine learning in decision-making processes. The remaining radiologists had never interacted with AI tools 
nor had they done any labeling for AI training. 
Radiologists attended a one-hour orientation session during which the lead
radiologist explained the purpose of the study (i.e., better understanding how to integrate AI in radiologists' decision-making processes), 
clarified the nature of the experimental task, and addressed questions and concerns. 
We assigned the radiologists
to two experimental groups, 
one for each workflow configuration,
balancing their years of experience and previous exposure to data labeling. 
In total, 19 participants started and completed the experimental task. 
The one- and two-step workflows had 11 and 8 submissions respectively, 
with five and three participants having prior exposure to data labeling for AI training.
The median years of experience (9) was identical across workflows.

\subsection{Measures and Statistical Analysis}\label{sec:measures}

\paragraph{Objective measures}
We assessed the impact of workflow configurations through the following measures:
\begin{itemize}
    \item Alignment between participants' diagnoses and AI inferences: Likelihood that the participants identify the same set of findings that are flagged by the AI.
    \item Diagnostic performance: Classification accuracy, false positive rate, false negative rate, and positive predicted values for both the AI and the participants' diagnoses. 
    \item Inter-rater reliability: Fleiss' kappa  for measuring agreement across participants \citep{fleiss1971measuring}.
    \item Time spent and confidence: Time spent reviewing each image, share of second opinions sought (a proxy for confidence), and likelihood estimates of the finding being present made by the participants. 
\end{itemize}
In a separate survey we conducted, participants reported that reviews of complex cases encountered on the job take between 10 and 20 minutes.
Thus, in the analysis of time on tasks we assumed that participants took breaks whenever they spent more than 15 minutes on a single image. Accordingly, we did not consider those observations (about 3\% of all cases) in the analysis. The arbitrary choice of this threshold (instead of, say, 10 or 20 minutes) does not affect our study findings. 
\paragraph{Taxonomy of findings}
The lead radiologist determined whether each of the 33 findings satisfied the following five non-mutually exclusive criteria:
(i) is critical, i.e., it requires immediate medical care and monitoring by healthcare professionals (vs. any other patient); 
(ii) is dangerous to overcall and treat if not actually present, i.e., it is important not to identify when absent; 
(iii) often requires a second opinion; 
(iv) has a vague definition, and (v) is common in animals and is often overlooked. 
We use these tags on findings later in Section~\ref{sec:results} to perform a disaggregated analysis for critical vs. non-critical findings, to investigate diagnostic performance on findings that are dangerous to overcall, and to clarify workflow effects for findings that are expected to have a high disagreement (i.e., majority vote in ground truth being less reliable) versus those where lower disagreement is expected (i.e., majority vote in ground truth being more reliable). To this end, we would expect high disagreement among radiologists in findings that are at least in two of the categories (iii), (iv), and (v). 

\paragraph{Subjective measures} 
We collected subjective measures of
participants' confidence, diagnostic performance, and trust in the tool. 
After each review, participants were asked whether the AI inferences helped them to make their diagnoses. 
In addition, the final questionnaire elicited answers on  seven-point Likert scales ranging from ``strongly disagree'' to ``strongly agree'' to assess the following measures:
\begin{itemize}
\item Workload: We inserted two questions related 
to the mental demand and
frustration dimensions from the NASA-TLX study \cite{hart1988development}.
\item Usefulness: We used two questions from the technology acceptance model (TAM)
of \citet{davis1989perceived} related to gains in speed 
and diagnostic performance obtained by using the AI tool.  
Participants also reported on changes in confidence working alongside the AI. 
In addition, they could describe instances where the AI was most useful and least useful during their decision making via free-text responses. 
\item Future use: Participants indicated whether they would use a similar AI diagnostic tool in their daily jobs 
and could elaborate on their preference in an open-ended question. 
\end{itemize}
To analyze differences in the ratings across workflows, 
we converted the Likert scale ratings into integers 1--7.

\paragraph{Statistical analysis}
Prior to data collection, we planned to study the
impact of workflow configurations on decision making 
by reporting summary statistics 
and the corresponding standard errors
for each workflow, e.g., point estimate\% [standard errors\%]. 
These standard errors 
(e.g., of positive predicted values)
are obtained via a nonparametric block bootstrap 
where the resampling is done at the participant's level, conditioning on their prior exposure to data labeling for AI training (see Section 3.8 of \citet{davison1997bootstrap}).
We test the null hypothesis of independence of 
outcomes for the participant-level summary statistics and workflow configurations via rank-sum permutation tests, again conditioning on prior exposure to data labeling \citep{hothorn2006lego}.
When relevant to the discussion, we report the corresponding one-sided (analysis of alignment between AI and participants) and two-sided (other analyses) p-values, considering a significance level of 0.1.
We also examine participants' reliance on AI flags by regressing their diagnoses
on AI confidence and a dummy variable that indicates the presence of the AI flag, interacted with workflow configuration. 
This approach is inspired by regression discontinuity designs, a methodology popular in the econometric literature  \citep{angrist2008mostly}. 
We fit the model via ordinary least squares and use sandwich standard errors clustered at the participant's level. 
Statistical significance of regression coefficients estimates is assessed via Wald tests.
The plan for the analyses of the aforementioned measures at the aggregate level and conditional on AI inferences was laid out before running the experiment and motivated our data collection efforts. 
Prior to data analysis, 
we conducted a power analysis 
relative to our investigation of anchoring effects. 
We estimated power to be about 60\%
for a 2\% standard deviation in the average agreement of participants with the AI and a difference of 1\% in participants' agreement with the AI between workflows. 
The analysis based on the findings taxonomy represents a post-hoc investigation motivated by our study findings, which we decided to report because it reveals valuable insights into the nature of anchoring effects. We conducted an additional analysis using generalized linear mixed models that account for prior participant exposure to data labeling. The results obtained through this methodology are similar to those described in Section \ref{sec:results} and thus are omitted.

\subsection{Experimental Platform}

The platform was developed using the Python-based framework Django. 
The platform can accommodate future studies in similar domains by enabling researchers to bring in their own data sets of images, lists of ground truth diagnoses, and AI flag thresholds on algorithmic confidence. Either of the workflow configurations can be used for such studies. The platform logs relevant data on a per image basis regarding human diagnostic decisions, time elapsed, and responses to subjective questions. The platform also allows individuals to implement comprehension checks and collect data via surveys after the diagnostic tasks are completed.

%% file: text/results.tex
\section{Results}
\label{sec:results}

\subsection{Alignment Between Participants and AI}\label{sec:anchoring}

\begin{mdframed}[style=MyFrame,  leftmargin=-\csname @totalleftmargin\endcsname, usetwoside=false]
\textbf{Result 1: 
Alignment between participants' diagnoses and AI inferences was highest in the one-step workflow, 
suggesting the influence of anchoring. This effect originated mostly from findings considered as non-critical for animal healthcare.}
\end{mdframed} 

\noindent The final diagnoses made by participants matched the AI flags
on 91\% [standard error=1\%] and 89\% [1\%] of the findings in the one- and two-step workflows, respectively. 
The alignment observed in the one-step workflow is significantly higher than in the two-step workflow (p-value is 0.04).
Note that, of all <image, finding> pairs, only 11\% were flagged as present by the AI, 
while 7\% of them were marked as present by majority vote in the dataset. 
The low prevalence is explained by the fact that animals usually present only a few (and luckily not most) of the 33 findings in our list. Thus, all results in this section and the differences between the two workflows need to be interpreted with this consideration in mind. 

\begin{figure*}
    \begin{center}
    \includegraphics[width=0.8\textwidth]{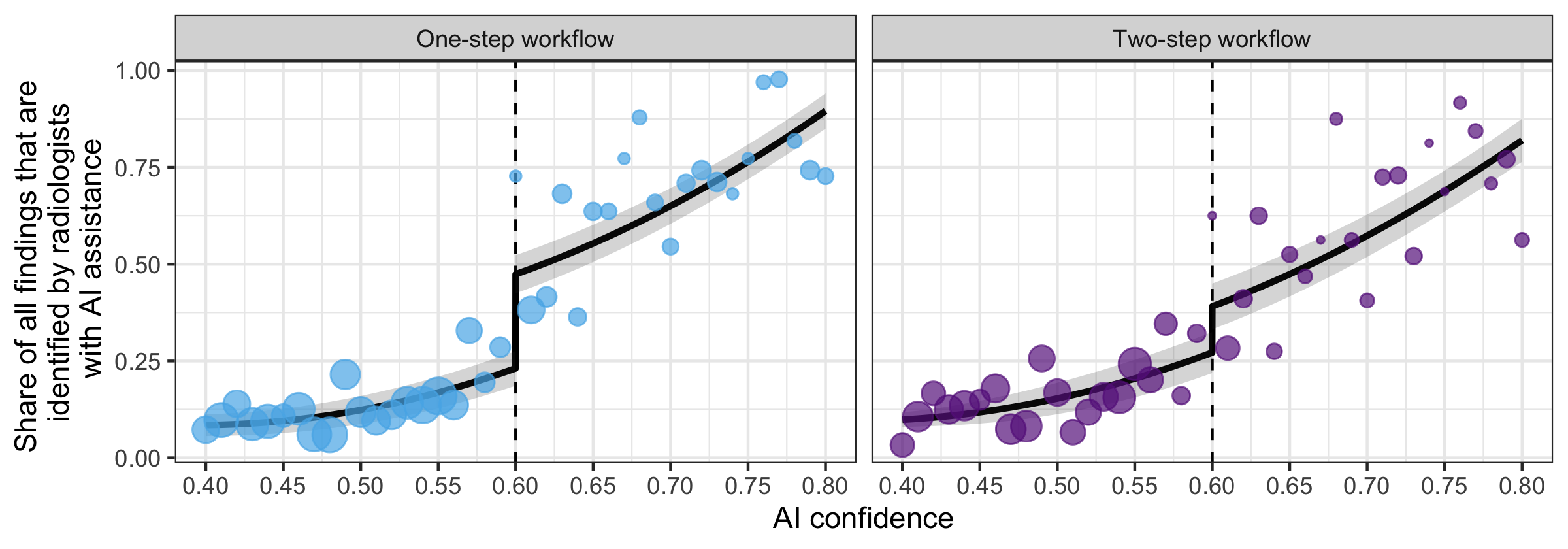}
    \end{center}
    \caption{
    Estimated probability that a participant would identify the radiographic finding in the X-ray as a function of AI confidence in the finding being present, for each workflow configuration. 
    Each dot represents the share of findings that were identified for all findings with a certain AI confidence, across the entire set of images. 
    The size of the dots is proportional to the number of findings. 
    The parametric fits and corresponding 90\% confidence intervals are represented by the solid black lines and gray shaded regions respectively. 
    The magnitude of the estimated discontinuity at 0.6 
    in the one-step workflow
    is substantially larger 
    than in the two-step workflow. 
    This finding indicates that the
    presence of AI flags, 
    which appeared for findings where AI confidence$\geq 0.6$ (vertical dashed line),
    had a stronger influence on diagnoses made by one-step workflow participants. Further analysis reveals that this impact originates mostly from non-critical findings. 
    }
    \label{fig:rd}
\end{figure*}

When an AI flag was present for findings, 
participants in the one-step workflow 
were significantly more likely to identify the finding as present in the X-ray
than their counterparts in the two-step workflow (in 71\% [3\%] and 65\% [3\%] of the findings, respectively; alignment is higher in the one-step workflow with p-value 0.07). 
When the AI flag was absent, participants 
in the one- and two-step workflows 
identified the finding in 7\% [1\%] and 8\% [1\%] 
of the cases (alignment is not significantly higher with p-value 0.18). 

To better understand the impact of AI and anchoring biases in each workflow, 
Figure \ref{fig:rd} shows the likelihood that participants identify a finding in the X-ray as a function of AI confidence.
We are interested in the comparison of 
the discontinuity in the likelihood at AI confidence=0.6, the value used to determine whether the AI flag is shown.
We observe that the estimated discontinuity 
for the participants' diagnoses in the one-step workflow is larger than 
for those in the two-step workflow 
(estimates of the dummy variable $\mathds{1}$(AI confidence$\geq$0.6) are 
0.24 [0.03] and 0.11 [0.03] 
in one- and two-step workflows, respectively;
difference is statistically significant with p-value$<$0.01).
This means that the presence of an AI flag substantially increased
the likelihood that participants
would identify the finding compared to the one-step workflow. 
A nonparametric analysis further corroborates this result: 
Findings with AI confidence between 0.6 and 0.65 were identified by participants 
48\% [3\%] and 40\% [4\%] of the times 
in one- and two-step workflows, respectively. 
Instead, 
those with AI confidence 
between 0.55
and 0.59
were identified 
20\% [3\%] and 25\% [4\%] of the times 
respectively. 
These results indicate the presence of anchoring effects on AI flags in the one-step workflow. 

A natural follow-up question is
whether participants in the one-step workflow 
relied more on the AI 
uniformly across all of the findings. 
Using the taxonomy described in Section \ref{sec:measures}, 
we can investigate the phenomenon
across two dimensions: 
the criticality of the finding
and the difficulty of its interpretation.
With respect to the criticality of findings 
(i.e., Question (i) in Section~\ref{sec:measures}),
we observe that anchoring effects were stronger 
on findings categorized as non critical.
For example, 
when a non-critical finding was flagged by the AI, 
participants identified it in the X-ray 
in 76\% [3\%] 
and 65\% [4\%] of the cases 
in one- and two-step workflows respectively. 
When a critical finding was flagged instead, 
the respective shares were 
67\% [3\%] and 65\% [3\%]. 
The discontinuity analysis
delivers similar results. 
One explanation of this result is that participants might have put more effort into making these diagnoses, while pondering less the possible presence of AI flags and potentially more the value of AI confidence. 
We also unsurprisingly observe that 
anchoring effects are salient 
on findings where disagreement among radiologists is expected to be largest (based on Questions (iii, iv, v) in Section~\ref{sec:measures}). 
When one of these finding was flagged by the AI, 
participants identified it in 73\% [4\%] and 64\% [4\%]
of the cases 
in one- and two-step workflows respectively. 
The gap is virtually zero
for the remaining not as difficult findings.

Finally, 
we briefly discuss 
whether and how 
participants in the two-step workflow 
revised their provisional diagnoses 
after observing AI inferences. 
In total, these participants changed their diagnoses on only 70 
of the 10560 findings evaluated. 
This corresponds to 5\% of all findings for which their initial diagnoses did not match the AI flags. 
The majority of these revisions (47)
occurred for findings 
that were flagged by the AI
but that the participants had not initially identified. 
Most of the remaining revisions (18)
also happened in cases of disagreement between AI and provisional diagnoses,
where the participant had initially identified the finding in the image
but the AI flag was absent. 
On the cases of initial disagreement, 
we could not detect any association between the tendency to revise and the criticality or difficulty of the finding.

\subsection{Diagnostic Performance}

\begin{table*}[t]
 \centering
 \footnotesize
  \caption{Diagnostic performance of AI alone and of participants' diagnoses made with AI assistance [standard error \%].}
  \resizebox{\textwidth}{!}{
  \begin{tabular}{l:l:ccccc}\toprule
    & & Accuracy
     & False Positive Rate & False Negative Rate
     & Positive Predicted Values
     & \% Predicted Positives \\  \midrule 
  \multirow{3}{*}{All findings} & AI & 94\% & 6\% & 18\% & 52\% & 11\% \\ 
  & One-step workflow & 91\% [1\%] & 8\% [1\%] & 16\% [3\%] & 42\% [3\%] & 14\% [1\%] \\ 
  & Two-step workflow & 90\% [1\%] & 9\% [1\%] & 17\% [3\%] & 39\% [3\%] & 14\% [1\%] \\ 
  \midrule
  \multirow{3}{*}{Critical findings}  & AI & 95\% & 5\% & 9\% & 55\% & 9\% \\ 
  & One-step workflow & 94\% [1\%] & 6\% [1\%] & 14\% [3\%] & 46\% [5\%] & 10\% [1\%] \\ 
  & Two-step workflow & 92\% [1\%] & 7\% [1\%] & 15\% [3\%] & 41\% [4\%] & 12\% [1\%] \\ 
  \midrule
\multirow{3}{*}{\shortstack[l]{Findings \\dangerous\\ to overcall}} & AI & 96\% & 4\% & 10\% & 51\% & 8\% \\ 
  & One-step workflow & 92\% [1\%] & 7\% [2\%] & 15\% [4\%] & 35\% [4\%] & 11\% [2\%] \\ 
  & Two-step workflow & 91\% [1\%] & 9\% [1\%] & 15\% [4\%] & 32\% [3\%] & 12\% [1\%] \\ 
  \midrule
  \multirow{4}{*}{\shortstack[l]{Findings with \\ lowest expected \\disagreement}} 
 & One-step workflow, AI correct & 96\% [1\%] & 4\% [1\%] & 11\% [2\%] & 56\% [4\%] & 8\% [1\%] \\ 
  & Two-step workflow, AI correct & 94\% [1\%] & 6\% [1\%] & 8\% [2\%] & 46\% [5\%] & 10\% [1\%] \\ 
  & One-step workflow, AI incorrect & 62\% [2\%] & 41\% [3\%] & 27\% [4\%] & 32\% [1\%] & 47\% [3\%] \\ 
  & Two-step workflow, AI incorrect & 66\% [4\%] & 36\% [4\%] & 25\% [7\%] & 35\% [3\%] & 44\% [4\%] \\
  \bottomrule
  \end{tabular}
  }
  \label{tab:diagnostic_performance}
\end{table*}
\begin{mdframed}[style=MyFrame,  leftmargin=-\csname @totalleftmargin\endcsname, usetwoside=false]
\textbf{Result 2: AI system outperformed participants across most of the  performance metrics considered. 
Participants in the one-step workflow
anchored more on the AI flags 
regardless of the AI accuracy, 
resulting in marginal gains in diagnostic performance when compared to the two-step workflow.
}
\end{mdframed} 

\noindent A critical dimension
related to the impact of workflow configurations on human decision making 
is diagnostic performance.
We conduct four related analyses of performance
that consider various characteristics of the findings (Table \ref{tab:diagnostic_performance}).
We now describe the key findings
from each of these investigations in turn. 

We start by considering 
all diagnoses made with AI assistance. 
The performance metrics relative to AI alone, 
one-step workflow, 
and two-step workflow participants
are reported in the first three rows 
of Table \ref{tab:diagnostic_performance}. 
We observe that 
the AI outperformed participants 
on most of the metrics. 
Nonetheless, 
the only notable---yet not statistically significant difference---in performance across workflows 
is in the positive predicted values 
(42\% vs. 39\% in one- and two-step workflows respectively). 
Classification accuracy
and false positive rate 
of the one-step workflow participants 
are also closer 
to those of the AI system, 
but 
the gains are minimal and 
our experiment 
is underpowered
to detect such small variations. 

Our second analysis 
focuses on critical findings. 
Similarly to the previous investigation,
we find that, 
while the AI outperformed both groups of participants,
those in the one-step workflow achieved 
slightly better performance, 
across all metrics.
We repeat the analysis 
on findings that 
may be dangerous to overcall, for which
making as few
false positive diagnoses
is crucial.
We observe that 
the AI achieved 
the lowest false positive rate (4\%),
followed by those of participants in the one-step 
and two-step workflows
(7\% and 9\% respectively). 


Evaluations of diagnostic performance 
can be inherently problematic: radiologists often disagree
on whether a certain finding 
is actually present, even in the original dataset.
We mentioned 
this phenomenon 
when describing 
the ground truth annotations
in Section \ref{sec:task}. 
Thus, 
for some of the findings,
a certain degree of disagreement between the diagnoses made by our study participants and ground truth should be expected.  
Our fourth analysis of performance 
focuses solely 
on the findings 
where we expect 
disagreement among radiologists 
to be lowest and thus ground truth annotations to be most reliable (again according to Questions (iii, iv, v) in Section~\ref{sec:measures}). 
We find that 
diagnoses made 
in the one-step workflow 
achieved higher accuracy and
lower false positive rates 
than those made in the two-step workflow. 
This mirrors our previous findings. 
However, given that the AI outperformed 
participants by a considerable margin, 
why don't we observe larger gains in performance? 
One explanation is that even the diagnoses made in the one-step workflow did not always match AI flags. 
Moreover, 
these participants tended to agree more with the AI flags even when they were inaccurate.
The last four rows of Table \ref{tab:diagnostic_performance}
report participants' diagnostic performance 
conditional on the accuracy of AI flags 
for these findings. 
We observe that, 
when AI advice was correct
(e.g., a finding that was present was flagged), 
participants in the one-step workflow
achieved better performance than those 
in the two-step workflow.
When AI advice was wrong
(e.g., a finding that was absent was flagged), 
they achieved worse performance, 
across all metrics. 
We observe an analogous phenomenon 
for critical findings with low expected disagreement, 
despite the minimal anchoring effects. 
These results demonstrate that
the stronger alignment between AI and participants  
observed in the one-step workflow 
was not always warranted: 
Showing the AI flags directly 
made participants 
more likely to identify the finding
not only when it was actually present 
but also also 
when ground truth indicated that it was not. 
These results explain the fact that anchoring 
led to only marginal gain
in participants' overall performance.


\subsection{Inter-rater Reliability}

\begin{mdframed}[style=MyFrame,  leftmargin=-\csname @totalleftmargin\endcsname, usetwoside=false]
\textbf{Result 3: 
Inter-rater reliability was highest for diagnoses made in the one-step workflow. 
}
\end{mdframed} 

\noindent We have mentioned 
at several points 
in the paper
that the diagnoses
in the ground truth annotations from individual radiologists
often differed. 
In our experiment, we expected
the presence of the AI to affect agreement among radiologists
differently across workflows. 
More specifically, 
we hypothesized that, as consequence of anchoring, 
(i) on findings where AI flags were accurate, agreement would be highest in the one-step workflow, i.e., one-step workflow participants would be more likely to make the same diagnoses;
and (ii) on findings where AI flags were most likely inaccurate,  
agreement would be lowest in the one-step workflow. 
For (ii), we consider the findings on which we expected low disagreement among radiologists as described in Section \ref{sec:measures}.
Thus, the hypothesized effects would run in different directions. 
We find that overall inter-rater reliability in the one-step workflow
is higher than in the two-step workflow, with the respective estimates of Fleiss' kappa being 0.55 and 0.49. 
For (i), 
the
inter-rater reliability 
measured on diagnoses 
made in the one-step workflow
is higher 
compared to those in the two-step workflow; 
the respective estimates of Fleiss' kappa
are 0.54 and 0.49. 
We find that the gap in kappas across workflows is largest
on findings that are considered as non-critical for animal healthcare, 
for which we also observe the largest anchoring effects.
We do not find 
evidence in support of (ii): 
Fleiss' kappas 
on findings 
with expected low disagreement are
0.37 and 0.35 
in one- and two-step workflows respectively.

\subsection{Time Spent on Decision Making and Confidence}\label{sec:time}

\begin{mdframed}[style=MyFrame,  leftmargin=-\csname @totalleftmargin\endcsname, usetwoside=false]
\textbf{Result 4: 
Time spent on the task did not differ across workflow configurations.
In cases of disagreement with the AI, 
one-step workflow participants sought more often second opinions than their counterparts. 
Evidence suggests that 
they 
might have weighed AI inferences 
more meaningfully.}
\end{mdframed} 
\begin{figure*}[t]
    \begin{center}
    \includegraphics[width=0.45\textwidth]{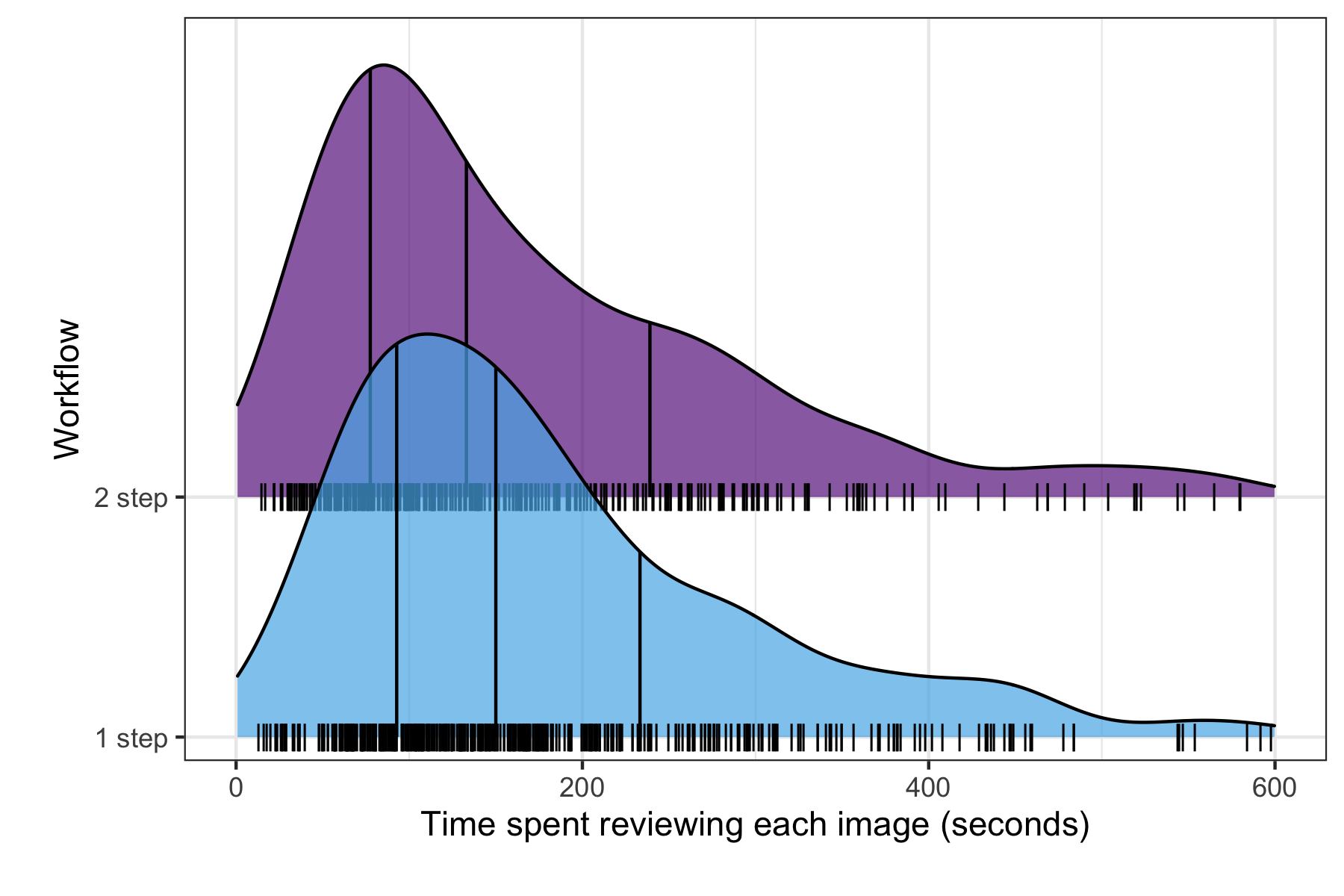}
    \includegraphics[width=0.45\textwidth]{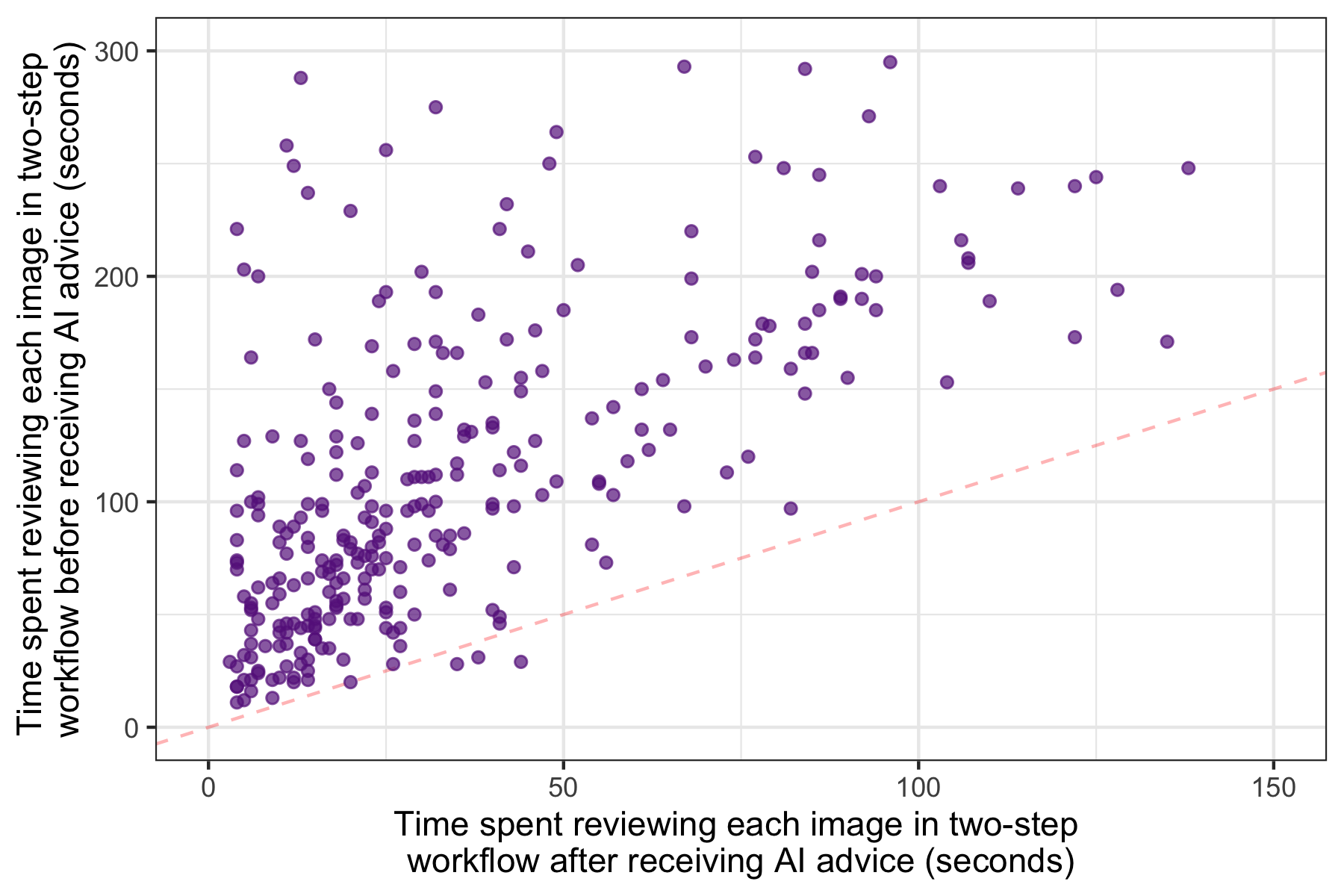}
    \end{center}
    \caption{Time spent by participants on decision making. 
    The plot on the left 
    shows the estimated densities of the time 
    spent by participants 
    reviewing the individual images
    and making the diagnoses, in each workflow.
    The vertical solid lines correspond 
    to first, second (median), and third quartiles. 
    The small vertical lines at the bottom 
    represent 
    individual observations, i.e., one image reviewed by one participant. 
    The plot on the right shows 
    the time participants in the two-step workflow 
    spent 
    before and after observing the AI inferences on each image
    (vertical and horizontal axes respectively). 
    If the time spent did not differ across phases, 
    the dots, which represent individual cases, 
    would lie around the 45 degree dashed red line.
    For visualization purposes, 
    we have limited the scales of the axes.
    }
    \label{fig:time}
\end{figure*}

\noindent We expected that requiring participants 
to make provisional diagnoses before
AI inferences were revealed
would substantially 
slow down their decision making. 
The distribution of the time 
participants spent 
reviewing the images and making the diagnoses 
for each workflow 
is shown 
in Figure \ref{fig:time} (left). 
We observe that 
participants 
in the two-step workflow
did not spend more time on the task
than those in 
the one-step workflow, 
neither in terms of 
the average nor of the median times
(the respective medians are 152 [standard error=20] and 139 [38] seconds, 
while averages are 189 [22] and 191 [36] seconds). 
We have identified 
two factors that
may explain 
this surprising result.
First, 
there exists a
large variability 
in the time spent 
by participants, which calls for a larger sample size. 
Indeed, 
the fastest participant
made the diagnoses
in an average time of 
slightly more than one minute, 
while reviews took over six minutes 
to the slowest participant. 
A second hypothesis is that one-step workflow 
participants might have considered AI inferences more carefully. 

We can investigate 
the second hypothesis 
by examining participants' 
need of second opinions 
across workflows. 
As a reminder, 
our study participants were asked whether, 
were they to encounter the
same patient in their daily job, 
they would seek the opinion of a colleague
before making the final call 
on the diagnosis. 
This option was rarely chosen 
and the overall rates of second opinions 
were similar across workflows 
(about 1\% of all findings evaluated). 
However, the two cohorts of participants tended to seek 
second opinions 
in different circumstances.
On the one hand, 
one-step workflow participants 
sought second opinions 
{\it more} often 
in cases where 
they {\it disagreed} with the AI inferences. 
For findings 
that were identified by participants 
but were not flagged by the AI,
the rates of second opinions 
were 
11\% [4\%] and 6\% [3\%]
for one- and two-step workflows respectively.
For findings that were flagged by the AI but 
were not identified by participants, 
the respective rates were 
2\% [1\%] and 1\% [1\%]. 
On the other hand, 
one-step workflow participants 
sought second opinions
{\it less} often
in cases where 
they {\it agreed} with AI inferences. 
This occurred in 
1\% [1\%] and 3\% [1\%]
of these findings 
that were flagged 
by both AI and participants
in one- and two-step workflows respectively.
The magnitude of these differences 
is substantially larger
in case of findings that are 
critical 
or difficult to interpret 
according to our taxonomy (Question (iii) in Section~\ref{sec:measures}). 
For example, 
for critical findings identified by the participant 
but not flagged by AI,
second opinions were sought in 
22\% [7\%] and 9\% [6\%]
of the cases in one-and two-step workflows respectively. 
However, the rates of second opinions 
largely differed across participants 
as some of the participants never sought second opinions at all 
(those with more years of experience did so less often).
Nonetheless, these empirical results appear to support the observation 
that participants in the one-step workflow
considered the AI advice more meaningfully, 
and varied the need of
second opinions 
according to their agreement
with the recommendations. 
At the same time, 
we show in Section~\ref{sec:anchoring} that
two-step workflow participants 
revised only a small number of their provisional diagnoses. 
Consistently, Figure~\ref{fig:time} (right) highlights 
that these participants often spent a small amount of time 
reviewing the AI assistance (horizontal axis), 
supporting a similar interpretation as the analysis of second opinions.

It is possible 
that variations in confidence 
were reflected
by the participants' subjective likelihood judgments of the findings being present. 
About three fourths of the collected estimates 
correspond to 0\%, 
which was the default value. 
In all these cases, 
it is possible that 
participants believed 
that the finding was certainly absent
or,
in the interest of time, 
that they didn't bother changing the default answer.
Consequently, we focus our analysis 
only on findings 
that participants identified in the images.
On these findings,
the investigation of second opinions
suggests that one-step workflow participants 
were less confident about their diagnoses compared to two-step workflow participants when the the AI flag was absent. 
However, we find that
the average likelihood estimates
for these findings are
comparable
across workflows (0.75 [0.02] and 0.77
[0.03] for one- and two-step workflows respectively, 
mean absolute differences 
between AI confidence 
and participants' estimates 
are 0.33 vs. 0.35). 
For findings that were instead flagged by the AI and were also identified by participants, 
two effects may
be at play. 
On the one hand, 
one-step workflow participants may 
have anchored on the AI confidence. 
On the other hand, 
the analysis of second opinions suggests 
that their
confidence may have been bolstered 
by the presence of the AI flag,
yielding likelihood estimates higher than the AI's. 
The average values of the likelihood estimates are similar across workflows
(0.89, [0.01] and 0.90 [0.02] 
in one- and
two-step workflows, respectively). 
The discrepancy 
between these estimates and the AI confidence 
(0.15 and 0.16 respectively) 
is also comparable across workflows. 
Thus, the analysis 
suggests that the magnitude of 
subjective likelihood estimates 
did not vary across workflows.

\subsection{Perceptions about AI inferences} 

\begin{mdframed}[style=MyFrame,  leftmargin=-\csname @totalleftmargin\endcsname, usetwoside=false]
\textbf{Result 5: 
Participants in the one-step workflow rated 
the AI advice as more useful. 
}
\end{mdframed} 
\begin{figure*}
    \begin{center}
    \includegraphics[width=\textwidth]{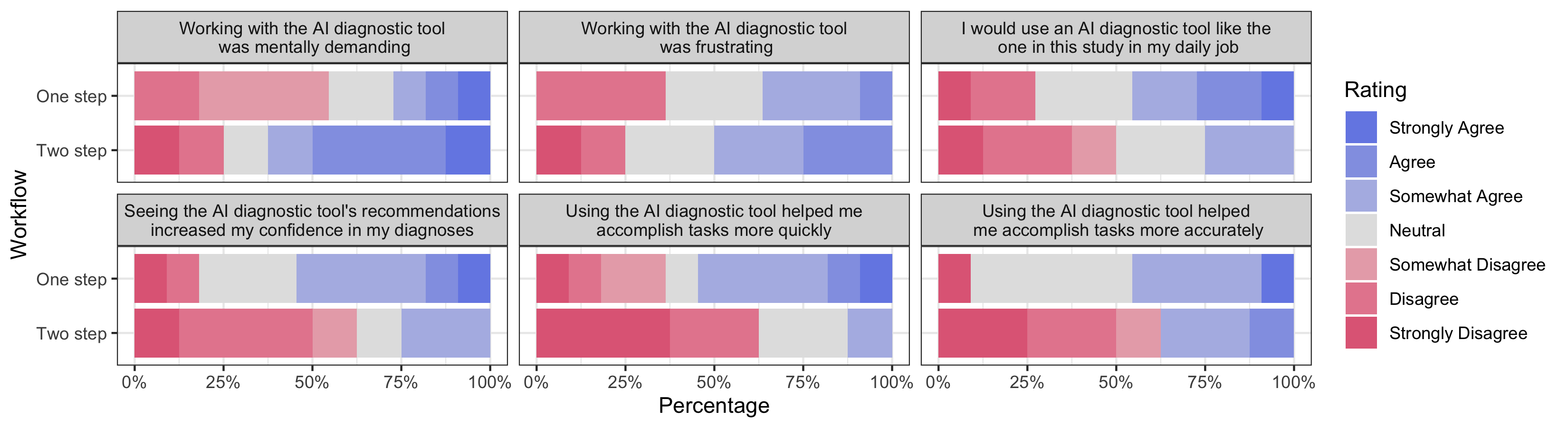}
    \end{center}
    \caption{Questions 
    and corresponding answers in the final survey elicited on a seven-point Likert scale.
    The questions, 
    reported in panel titles, 
    concern perceived workload, 
    future use of the AI tool, 
    and perceived utility of the AI inferences.
    }
    \label{fig:lklratings}
\end{figure*}
\noindent After reviewing each image and making their diagnoses,
participants were asked 
whether the AI advice had been useful. 
Those assigned to the one-step workflow
reported that the AI was useful 
for 36\% [standard error=7\%] of the images,
whereas 
the two-step workflow participants 
found it useful in only 17\% [4\%] of the cases 
(p-value for hypothesis of independence is 0.11). 
The difference across workflows
in the share of instances 
where the AI help was deemed useful 
is less than 10\% for the first five images 
shown in the experiment
and as large as 30\% for the last five. 
Three questions 
in the final questionnaire 
can help us disentangle the AI's utility, 
or the lack thereof, 
into gains in accuracy,
speed, and confidence in the diagnoses. 
Figure \ref{fig:lklratings} shows
the distribution of these ratings.
We observe that participants 
assigned to the two-step workflow 
gave substantially lower ratings 
(or equivalently less utility)
across all three dimensions. 
For gains in accuracy, 
the average ratings 
on the converted 1--7 Likert scale 
are 4.4  and 3.1 
for one- and two-step workflows respectively. 
The respective ratings for confidence 
are 4.3  and 3.0  respectively. 
The difference relative 
to gains in speed 
is particularly striking: 
The average rating is 4.2 for one-step workflow 
participants
and only 2.5 for those in the two-step workflow. 
This indicates that two-step workflow participants felt that the AI slowed down their decision making more than those in the one-step workflow. 
Two-sided p-values for tests of independence relative to confidence and speed are 0.1 and 0.05 respectively, while the p-value relative to accuracy is 0.26. 
In line with these results,
we observe that 
two-step workflow participants 
reported the task to be slightly
more stressful and the workload to be more demanding 
than their counterparts 
(average aggregate scores are 3.8  
and 4.4  for one- and two-step workflows, 
with lower means indicating less stress and demand; 
however, we do not reject the null hypothesis of independence).

When asked for which diagnoses the AI advice 
had been {\it most} helpful, 
most of the participants 
(among those who answered) 
indicated 
that the AI helped them identify minor or incidental findings 
that would be less important in terms of decision making to the evaluating radiologist. 
This observation is consistent with the results of the quantitative analyses in Section~\ref{sec:anchoring} that highlighted 
more alignment between participants and AI inferences for non-critical findings.
Participants also 
responded that 
the AI tool ``was very helpful to reinforce [their] confidence'',  
could be used when they were ``on the fence about a finding instead of asking a colleague their opinion'',
and made them second guess their opinions by asking themselves whether 
they ``could `see' why it may have read it that way''.
When asked about the diagnoses for which the AI advice 
had been {\it least} helpful,
participants  
indicated that 
for findings are erroneously 
flagged by the AI tool 
``it's an extra step to `ignore' it'' 
and that they spent ``time searching for something that [they] ultimately decided isn't there''. 
We did not identify notable 
differences in the answers across workflows. 

Two-step workflow participants also appeared to be less willing
to use this AI tool 
in their daily jobs
(average ratings for one- and two-step workflows 
were 4.2 and 3.2 respectively), 
although we cannot reject the null hypothesis of independence. 
The participants who were reluctant to use the tool in the future
expressed their frustration 
with the fact that the tool 
did not smoothly integrate 
into their workflows.
Some reported that the AI tool
was often (in their opinion) inaccurate and nudged them to spend extra time evaluating certain findings. This echoes the results described in the previous paragraph.
One participant argued that, while the utility of the AI in its current form appeared to be limited, ``if the tool were able to correctly interpret the images as [they] would and incorporate those findings into a report that [they] could then edit this could increase productivity''. 

%% file: text/discussion.tex
\section{Discussion}
\label{sec:newdiscussion}
\noindent \paragraph{Task realism} 
Several limitations need to be considered while interpreting the results of this study. First, the task setup did not include background clinical information on the patient (e.g., notes or historical background on the animal),
which is typically available to the radiologist. 
Second, radiologists generally have access to multiple X-ray images and views from the patient. In our study, only one image was provided. 
The choice was motivated by the fact that the AI is only trained on individual images and does not leverage other clinical information. Having more views and clinical information available, radiologists may have exhibited different behaviors.
Third, it is possible that some participants might pay more attention in their daily assessments than they did in our experiment given that they knew that these decisions would not impact animal treatment. 
Fourth, our study included only a short onboarding process covering the task and description of the AI tool. In real-world deployments, this phase should ideally be longer and provide more detailed information on the tool and its intended use \citep{jacobs2021designing, cai2019hello, sendak2020human}. 

\noindent \paragraph{Choices of interaction design} 
The platform entailed a series of design choices that require further analysis in future studies as they may have important effects in the interaction design of the workflow.
Were the UI to change, for example through the removal of the AI flags from the navigation bar, our results could be affected. 
Examples of alternative designs include participants being asked to evaluate only a few of the findings at a time or review only the cases of disagreement with the AI inferences perhaps at a later stage. Further modifications of the interface could involve a dynamic selection of the workflow depending on findings based on the AI confidence.
For instance, we observed that the agreement between participants and AI flags increased with AI confidence. 
Thus, it may be preferable for radiologists to first investigate findings where the uncertainty of AI inferences is highest. 
This proposal partially aligns with approaches explored by prior work, where human review was required for the assessment of the most difficult tasks \citep{raghu2019algorithmic, madras2017predict}. 
Our findings also suggest that adjusting the threshold used to set AI flags by finding type (0.6 in our case) can influence final diagnostic decisions.   
Additionally, our participants used a novel experimental platform over two task sessions. Learning effects may have also impacted our experiment, particularly in terms of trust, ease of use, and time-related analyses. Such effects might have been more visible if the experiment had included more sessions.

\noindent \paragraph{Appropriate reliance} This study frames a question about the possibility of developing designs that could provide the best of both workflows: How might we obtain higher user satisfaction and engagement with the AI inferences seen in the one-step process while avoiding the increased tendency to anchor on erroneous AI inferences? 
We believe there is promise in studying modifications of the one-step workflow.
For example, similarly to judges who deviate from sentencing guidelines per special considerations of the situation \citep{parai2020},
radiologists could be asked to write down the reasons behind their diagnoses in case of disagreement with the AI inferences on a critical finding where AI confidence is far from the decision boundary.
Alternatively,
the opinion of a second radiologist may be required. 
This is, for example, what occurs in child welfare hotline screening decisions, where calls regarding children at high risk of out-of-home placement can only be screened out pending supervisor's approval \citep{de2020case}. 
At the same time, these potential modifications must be balanced with prior findings that cognitive forcing functions decrease trust and willingness to work with AI assistants \citep{buccinca2021trust}. 
Another possible strategy for trust calibration is represented by model explanations, beyond the likelihood estimates presented as AI confidence in our work. 
This direction has been explored by past work and holds potential in the clinical imaging context \citep{diprose2020physician, wang2019designing}.

\section{Conclusions}
\label{sec:newconclusion}
We evaluated two methods for integrating AI inferences about radiographic findings into the workflows of veterinary radiologists, seeking to understand how the different approaches influence decision making. 
Our findings revealed that radiologists' diagnoses were more aligned with AI advice when it was shown immediately than in workflows where AI inferences were displayed after the radiologist had rendered a provisional assessment. 
The alignment, however, was similar across workflows for findings that were considered to be critical for the animal. 
Diagnoses made in the one-step workflow were characterized by marginal gains 
in diagnostic performance and higher levels of inter-rater reliability compared to those in the two-step workflow.
Radiologists in the one-step workflow more frequently sought second opinions in cases of disagreement with the AI than in the two-step workflow 
and rated the AI tool was rated as more useful. 

These results suggest that the one-step workflow can be meshed more smoothly with
the current decision-making processes of radiologists than the two-step workflow. 
The dissatisfaction with AI assistance observed in the two-step workflow 
may be explained by the costs of task switching, interruption, 
and recovery described in Section \ref{sec:background}. Adding a second step requires radiologists to stop, reassess, 
and reaffirm the diagnoses they had just completed.
As a result, the AI advice shown after an analysis 
by the radiologists appears more prone to being disregarded.
However, alignment between AI and radiologists was stronger in the one-step workflow even when AI inferences were inaccurate, suggesting increases in inappropriate reliance and anchoring on AI inferences.


At the highest level, our experiment demonstrates the importance of interaction design for clinical AI systems. 
We have explored a fundamental dimension of human-AI workflow, considering the effects of whether the AI inferences are made available immediately or following an AI-free analysis. Much remains to be explored.

\begin{acks}
This study would not have been possible without the contribution and expertise of the veterinarian radiologists at Antech Imaging Services. The authors also thank Paul Koch for his help with the management of the infrastructure for the experiment, Lisa Ziemer for her support, and the anonymous reviewers for their valuable suggestions. 
\end{acks}

%% file: text/conclusions.tex
\label{sec:conclusions}